\documentclass[english,preprint,aps,nofootinbib, 11pt]{revtex4}
\usepackage[T1]{fontenc}
\usepackage[latin1]{inputenc}
\usepackage{amsmath}
\usepackage{graphicx}
\usepackage{amssymb}
\usepackage{graphics}
\usepackage{babel}

\makeatletter

\newcommand{\bc}{\begin{center}}
\newcommand{\ec}{\end{center}}

\makeatother

\begin{document}

\title{Geometro-thermodynamics of tidal charged black holes}
\author{L\'aszl\'o \'Arp\'ad Gergely$^{1,2}$, Narit Pidokrajt$^{3}$, Sergei Winitzki$%
^{4}$}
\affiliation{$^{1}$ Department of Theoretical Physics, University of Szeged, Tisza Lajos
krt 84-86, Szeged 6720, Hungary\\
$^{2}$ Department of Experimental Physics, University of Szeged, D\'om T\'er 9,
Szeged 6720, Hungary\\
$^{3}$ Department of Physics, Stockholm University, 106 91 Stockholm, Sweden%
\\
$^{4}$ Arnold Sommerfeld Center for Theoretical Physics, Department of
Physics, Ludwig-Maximilians University, Theresienstr. 37, 80333 Munich,
Germany }

\begin{abstract}
Tidal charged spherically symmetric vacuum brane black holes are
characterized by their mass $m$ and tidal charge $q$, an imprint of the
5-dimensional Weyl curvature. For $q>0$ they are formally identical to the
Reissner-Nordstr\"om black hole of general relativity. We study the
thermodynamics and thermodynamic geometries of \ tidal charged black holes
and discuss similarities and differences as compared to the Reissner-Nordstr\"om black hole. As a similarity, we show that (for $q>0$) the heat capacity of
the tidal charged black hole diverges on a set of measure zero of the
parameter space, nevertheless both the regularity of the Ruppeiner metric
and a Poincar\'e stability analysis shows no phase transition at those points.
The thermodynamic state spaces being different indicates that the underlying
statistical models could be different. We find that the $q<0$ parameter
range, which enhances the localization of gravity on the brane, is
thermodynamically preferred. Finally we constrain for the first time the
possible range of the tidal charge from the thermodynamic limit on
gravitational radiation efficiency at black hole mergers.

\end{abstract}

\date{\today }
\maketitle

\section{Introduction}

Thermodynamics of black holes (BHs) was formulated almost 40 years ago by
Bardeen, Carter and Hawking \cite{bh-thermodynamics}. The Bekenstein-Hawking
entropy is proportional to the area of the event horizon and it reads 
\begin{equation}
S=\frac{c^{3}k_{B}A}{4G\hbar }~,  \label{entropydef}
\end{equation}%
where $c,k_{B},G$ and $\hbar $ are speed of light, Boltzmann's constant,
Newton's gravitational constant and the reduced Planck's constant,
respectively and $A$ represents the area of the horizon.

Despite the fact that BH thermodynamics is well established, there is no
controlled calculation of BH entropy based on standard statistical mechanics
which associates entropy with a large number of microstates. The first paper
on the BH's microstate counting appeared in 1996 by Strominger and Vafa \cite%
{Strominger:1996sh} who were able to calculate the Bekenstein-Hawking
entropy of a five-dimensional extremal BH in the framework of string theory.
Since then there have been a growing number of papers on this topic.

An alternative way to study BH thermodynamics phenomenologically is by
studying a certain pseudo-Riemannian geometry defined on the thermodynamic
state space (the parameter space). A {}\textquotedblleft
thermodynamical\textquotedblright\ metric known as the Ruppeiner metric is
defined as the negative of the Hessian of the entropy $S$ with respect to
the mass $m$ and other extensive parameters $q^{i}$, including electric
charge and angular momentum: 
\begin{equation}
g_{ab}^{R}=-\partial _{a}\partial _{b}S(m,q^{i})~.  \label{Ruppeiner}
\end{equation}%
The first paper on the subject is by Ruppeiner~\cite{first-ruppeiner}; a
review on the application to various thermodynamic systems is given in~\cite%
{Ruppeiner:1995zz}. The Ruppeiner geometry of anyon gas has also been worked
out~\cite{anyon}.

The Ruppeiner geometry is one particular type of information geometry~\cite%
{Johnston:2003ed}. Ruppeiner originally developed his theory in the context
of thermodynamic fluctuation theory, for systems in canonical ensembles.
Many BHs have \textit{negative} specific heats and are described
microcanonically. For systems with non-interacting underlying statistical
models, such as the ideal gas, the Ruppeiner geometry is flat~\cite%
{first-ruppeiner}. Singularities in the curvature of the Ruppeiner metric
signal thermodynamic instabilities of the system in question.

The most physically significant result as reported in~\cite{Aman:2005xk} is
that the Myers-Perry Kerr BH ultraspinning instability is hinted by the
Ruppeiner geometry. More precisely, the curvature singularities of the
Ruppeiner metric signal the onset of ultraspinning instabilities of the
Myers-Perry Kerr BHs in dimensions $D>5$, found earlier by Emparan and Myers 
\cite{Emparan:2003sy}.

In~\cite{Aman:2006kd} it was shown that the Ruppeiner metric for a
two-dimensional thermodynamic state space is flat, provided the entropy
takes the power-law form $S=m^{p}f(Q/m)$ for any $p\neq 1$, where $Q$
represents a conserved charge (the angular momentum in the case of the
spinning Kerr BH). This theorem is useful when one wants to quickly rule out
thermodynamic curvature singularites of the system. Certain BHs do have flat
Ruppeiner metric, \textit{e.g.\ } Reissner-Nordstr\"om (RN), BTZ, and
4-dimensional Einstein-Maxwell-dilaton BHs. Well known examples of BHs with
non-flat Ruppeiner geometry include Kerr, Kerr-Newman, and Reissner-Nordstr\"om-AdS BHs~\cite{Aman:2003ug}. Recent papers on applications of the Ruppeiner
geometry to BH thermodynamics are listed in \cite{papers}-\cite{papers1}.
Thermodynamic geometry has been accepted as one of the standard tools used
in investigating whether instability or critical phenomena is present in a
given thermodynamic sytem.

An alternative geometric approach to thermodynamics is given by the Weinhold
metric~\cite{first-weinhold}, which is defined as the Hessian of the mass
(internal energy) with respect to entropy and other extensive parameters
such as charge, angular momentum, etc. 
\begin{equation}
g_{ab}^{W}=\partial _{a}\partial _{b}m(S,q^{i})~.  \label{Weinhold}
\end{equation}%
The Weinhold metric is equal to the conformally rescaled Ruppeiner metric,%
\begin{equation}
g^{W}=Tg^{R}~,  \label{conformal}
\end{equation}%
where it is understood that both metrics have been transformed to the same
set of coordinates. The conformal factor $T$ is the temperature,%
\begin{equation}
T=\partial _{S}m=\frac{1}{\partial _{m}S}~.  \label{temperature}
\end{equation}%
The relationship~(\ref{conformal}) has been derived previously using
involved thermodynamical arguments~\cite{Salamon1984}. We present in
Appendix \ref{sec:geodesics} a generic and simple proof of the statement~(%
\ref{conformal}).

Black hole solutions arise not only in general relativity and string theory,
but also in brane-world gravity models. There are many brane-world
scenarios, but in the simplest gravity evolves in a curved 5D space-time
(the bulk), which contains a temporal 4D hypersurface (the brane), on which
all the field of the standard model are localized. Gravitational dynamics on
the brane is governed by an effective Einstein equation \cite{SMS}-\cite%
{VarBraneTension}.

The most well-known brane BH is the spherically symmetric vacuum \textit{%
tidal} charged BH, derived in \cite{tidalRN}: 
\begin{equation}
ds^{2}=-f\left(r\right)dt^{2}+f^{-1}\left(r\right)dr^{2}+r^{2}\left(d%
\theta^{2}+\sin^{2}\theta d\varphi^{2}\right)\,.  \label{RN}
\end{equation}
The metric function $f$ is given as 
\begin{equation}
f\left(r\right)=1-\frac{2m}{r}+\frac{q}{r^{2}}\,.
\end{equation}
Such BHs are characterized by two parameters: their mass $m$ and tidal
charge $q$. The latter arises from the Weyl curvature of the 5D space-time
into which the brane is embedded (more exactly, from its {}``electric'' part
as computed with respect to the brane normal).

Formally the metric (\ref{RN}) agrees with the Reissner-Nordstr\"om solution
of a spherically symmetric Einstein-Maxwell system in general relativity,
provided we replace the tidal charge $q$ by the square of the \textit{%
electric} charge $Q$. Thus $q=Q^{2}$ is always positive, when the metric (%
\ref{RN}) describes the spherically symmetric exterior of an electrically
charged object in general relativity. By contrast, in brane-world theories
the metric (\ref{RN}) allows for any sign of $q$. A positive tidal charge
weakens the gravitational field of the BH in precisely the same way the
electric charge of the Reissner-Nordstr\"om BH does. A negative tidal charge,
however, strengthens the gravitational field, contributing to the
localization of gravity on the brane.

The structure of the tidal charged BH in the case $q>0$ is in full analogy
with the general relativistic Reissner-Nordstr\"om solution\footnote{%
In making analogy with the RN BH one can also consider the Born-Infeld BHs,
which is a nonlinear generalization of the RN BH~\cite{Myung:2008eb}}. For $%
0<q<m^{2}$ it describes tidal charged BHs with two horizons, located at $%
r_{\pm }=m\pm \sqrt{m^{2}-q}$, both below the Schwarzschild radius. For $%
q=m^{2}$ the two horizons coincide at $r_{e}=m$ (this is the analogue of the
extremal Reissner-Nordstr\"om BH). For any $q<0$ there is only one horizon, at 
$r_{+}=m+\sqrt{m^{2}+\left\vert q\right\vert }$. For these BHs, gravity is
increased on the brane by the presence of the tidal charge. This contributes
towards the localization of gravity on the brane.

Although the full 5-dimensional solution containing the tidal charged BH as
the brane section remains unknown, very recently it has been proven at a
perturbative level, that when the tidal charged contribution dominates over
the Schwarzschild contribution, the horizon does close in the fifth
dimension, therefore the tidal charged brane BH becomes a section of a
5-dimensional BH with regular horizon~\cite{new1} (see also the related Ref.~%
\cite{new2}).

Work on the tidal charged BH includes the matching with an interior stellar
solution, a procedure requiring a negative $q$ \cite{GM}, the study of weak
deflection of light to second order in both parameters \cite{GeDa}, \cite%
{eikonal}, the study of weak gravitational lensing by tidal charged BH \cite%
{tidalLens}, a confrontation with solar system tests \cite{BohmerHarkoLobo},
and the evolution of thin accretion disks in this geometry \cite{accretion}.

In this paper we analyze the tidal charged BH both by standard thermodynamic
and geometric methods provided by the thermodynamic metrics. Emparan,
Horowitz and Myers have constructed a lower-dimensional toy model with BTZ
BHs on the brane, which are BTZ {}\textquotedblleft black
strings\textquotedblright\ in the bulk \cite{EHM}. They were able to show
that the four-dimensional entropy computed from the horizon area agrees (to
a leading order at large mass) with the three-dimensional entropy computed
from the circumference of the horizon. The two definitions of the entropy
differed in the number of dimensions and in the value of the corresponding
Planck masses. Therefore the result of~\cite{EHM} can be taken as a hint
that there are thermodynamic constraints on the 5D extension of the tidal
charged BH, which are provided by our present analysis.

In Section~\ref{sec:Thermodynamic-considerations} we discuss the mass and
tidal charge dependence of the entropy of the tidal charged BH. We compute
the temperature and analyze the heat capacity at constant $q$. Then in
Section~\ref{sec:Thermodynamic-geometry} we study in detail the Ruppeiner
and Weinhold geometries of the tidal charged BH, pointing out similarities
and differences with respect to the Reissner-Nordstr\"om BH. The geodisic
structure for both information geometries is presented in Appendix.

One major concern is the stability of any BH solution. This can be
investigated by thermodynamics means. In this context it has been shown that
for charged AdS BHs the plot of $1/T$ vs. horizon area is quite similar to
the ($p,V$) diagram of the Van der Waals gas, indicating phase transitions 
\cite{Ch-prd}. Based on our analysis we will prove in Section \ref%
{sec:stable} that the tidal charged BHs are stable, regardless of the sign
of the tidal charge. Then, in Section \ref{sec:limit} we restrict for the
first time the range of the tidal charge, based on thermodynamic limits on
gravitational radiation efficiency at BH mergers.

Finally we discuss our results in the concluding section.

\section{Thermodynamic considerations\label{sec:Thermodynamic-considerations}%
}

\subsection{Entropy and mass}

\noindent The exterior horizon for $q>0$ or the single horizon for $q<0$ are
both given by 
\begin{equation}
r_{+}=m+\Theta ~,
\end{equation}%
where we have introduced the shorthand notation $\Theta =\sqrt{m^{2}-q}$,
real for any $q\leq m^{2}$. The BH's entropy (\ref{entropydef}) in
geometrized units and $k_{B}=1/\pi $ reads, 
\begin{equation}
S=\frac{A}{4\pi }=r_{+}^{2}=\left( m+\Theta \right) ^{2}~.  \label{entropy}
\end{equation}

\begin{figure}[tbp]
\includegraphics[width=9.5cm]{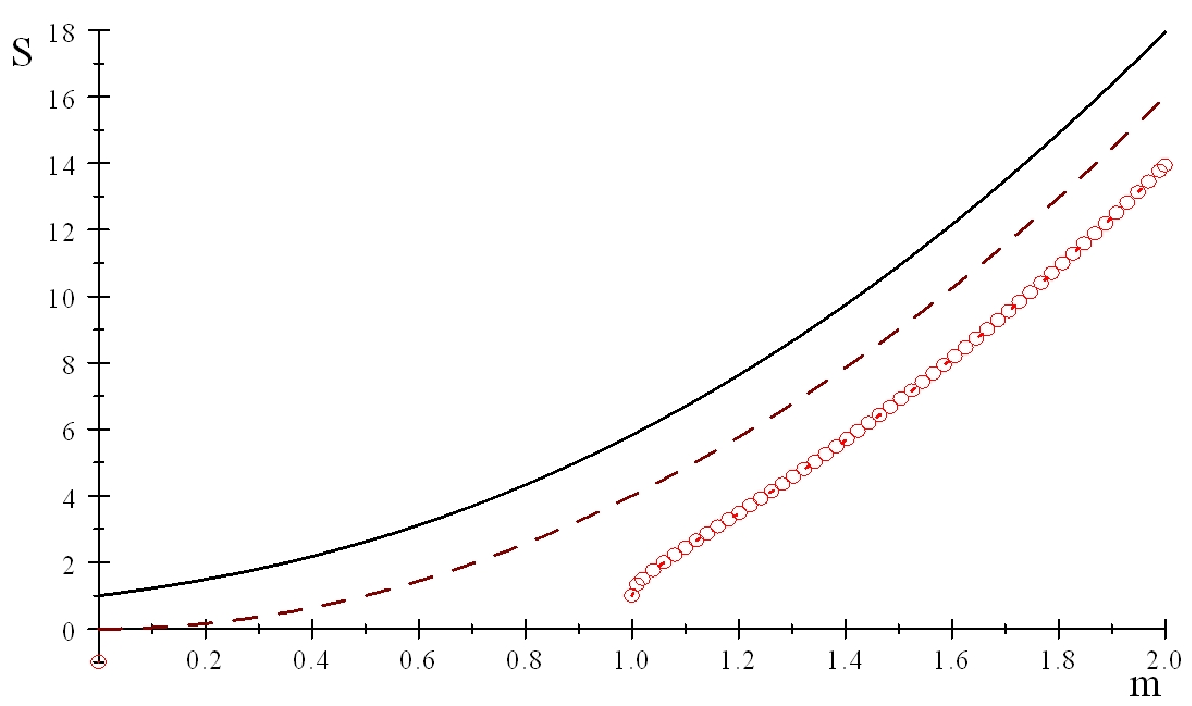} 
\caption{The entropy always increases with the mass. The plots are for
negative tidal charged BH with $q=-1$ (solid line), Schwarzschild BH with $%
q=0$ (dashed line) and positive charged tidal BH with $q=1$ (dotted line).
In the latter case one can see that the entropy is not defined for masses
below the extremal limit $m<\protect\sqrt{q}$.}
\label{Fig1}
\end{figure}

\begin{figure}[tbp]
\includegraphics[width=9.5cm]{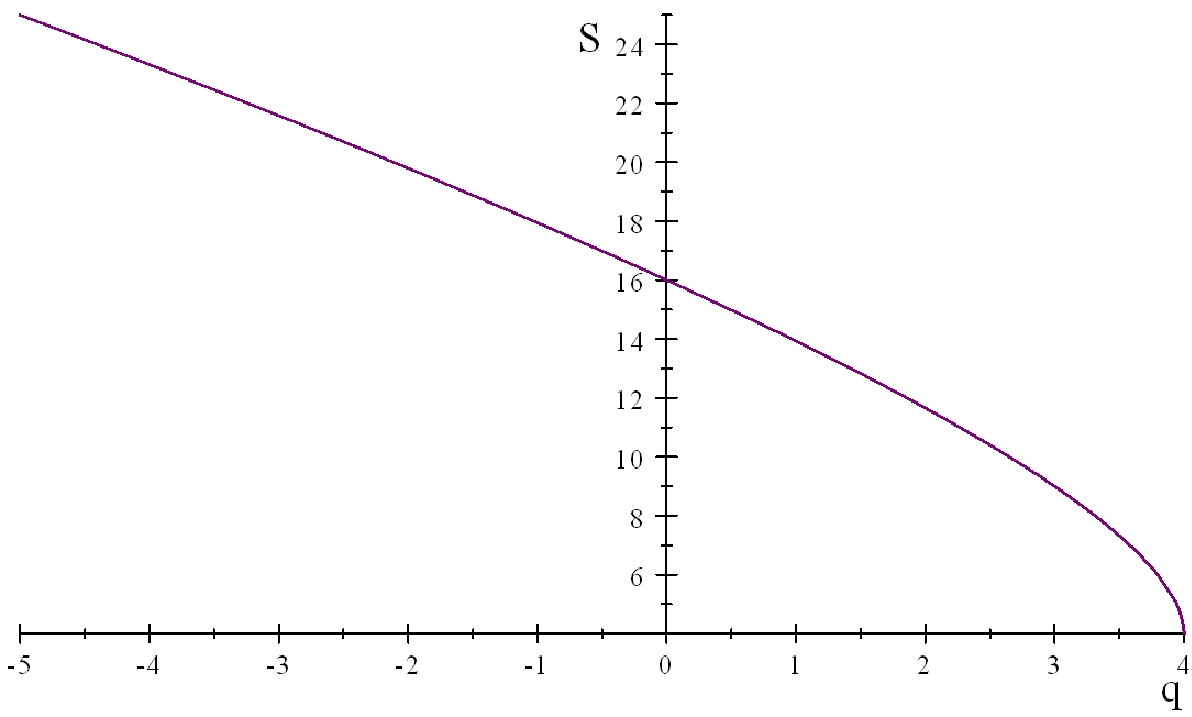} 
\caption{The qualitative behaviour of entropy vs. tidal charge. The entropy
decreases with increasing $q$. The plot is for $m=2$.}
\label{Fig2}
\end{figure}

It is instructive to see the variation of entropy with the BH parameters.
The entropy increases with mass (Fig.~\ref{Fig1}). By contrast, the entropy
decreases with increasing $q$ (irrespective of the sign of $q$) until $%
S=m^{2}$ and the minimal horizon area $A=4\pi m^{2}$ are reached at the
extremal limit $q=m^{2}$ (Fig.~\ref{Fig2}). This is to be expected since for 
$q>m^{2}$ the metric (\ref{RN}) describes a naked singularity (thus the area 
$A$ is undefined).

We can also express the mass of the tidal charged BH in terms of entropy and
tidal charge: 
\begin{equation}
m=\frac{\sqrt{S}}{2}\left( 1+\frac{q}{S}\right) .
\end{equation}%
The first law of thermodynamics is 
\begin{equation}
dm=TdS+\psi dq
\end{equation}%
where 
\begin{equation}
\psi =\frac{\partial m}{\partial q}=\frac{1}{2\sqrt{S}}
\end{equation}%
is the potential associated with the tidal charge.

Although the solution (\ref{RN}) is static, in a classical quasi-stationary
process the mass has to be either conserved or to slowly increase\footnote{%
For now, we disregard the Hawking radiation, which is negligible for
astrophysical black holes.}, in order to obey the second law of
thermodynamics. This could happen by an accretion process. Similarly, the 5D
geometry may evolve in a quasi-stationary way only such that the tidal
charge is conserved or it decreases. 

\subsection{Hawking temperature and heat capacity}

By the first law of thermodynamics, the Hawking temperature of the BH is
given by 
\begin{equation}
T(m,q)=\partial_{S}m=\frac{1}{\partial_{m}S}=\frac{\Theta}{%
2\left(m+\Theta\right)^{2}}~.  \label{T}
\end{equation}
The same value $T(m,q)$ is found by computing the temperature of the Hawking
radiation if one uses the well-known formula for the surface gravity of a
spherically symmetric Killing horizon (see e.g.~\cite{Padmanabhan:2003gd}).

The temperature $T(m,q)$ increases with $q$ for $q<0$ up to the maximal
value $T=1/(8m)$ at $q=0$, then decreases with increasing $q>0$ down to $T=0$%
, which is reached in the extremal limit ($q=m^{2}$). Thus the minimal
entropy belongs to $T=0$ and the hottest BH with a given mass is for $q=0$,
representing the Schwarzschild BH.

\begin{figure}[tbp]
\includegraphics[width=9.5cm]{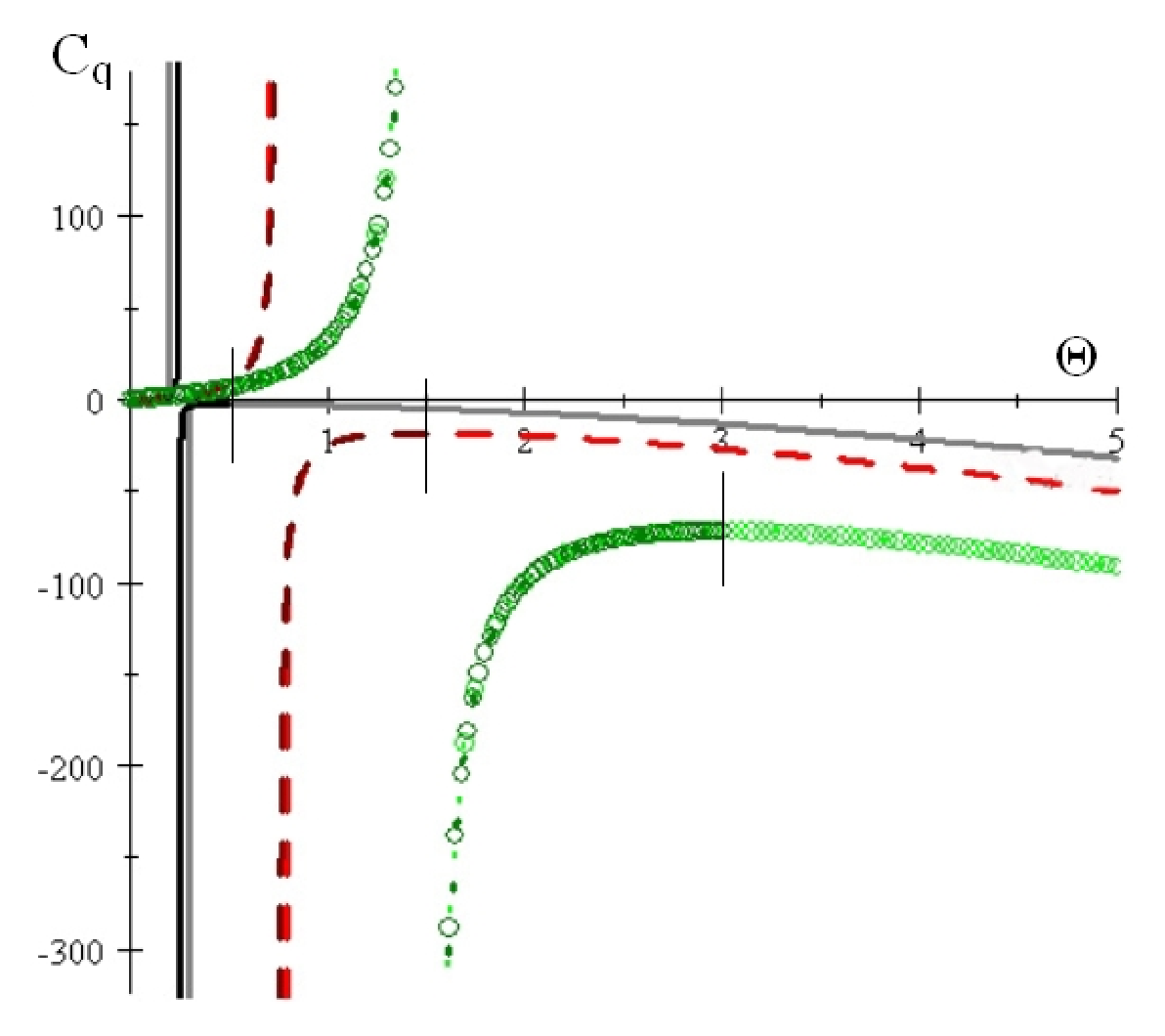} 
\caption{The heat capacity $C_{q}$ as function of $\Theta $ for $m=0.5$
(solid curve), $m=1.5$ (dashed curve) and $m=3$ (dotted curve). All these
curves show a vanishing heat capacity in the extremal case (at $\Theta =0$)
and a divergent behaviour at $\Theta =m/2$ (equivalent to $q=3m^{2}/4$ or $%
S=3q$). Up to the Schwarzschild limit (at $\Theta =m$, marked with a bar on
the graphs) the curves apply both for tidal-charged black holes and for
Reissner-Norstr\"om black holes (with electric charge $Q=\protect\sqrt{q}$).
The region of the curves for $\Theta >m$ (thus $q<0$) applies only for
tidal-charged black holes. }
\label{Fig3}
\end{figure}

The heat capacity of this BH at constant $q$ is given by 
\begin{equation}
C_{q}=\frac{\partial m}{\partial T}=T\frac{\partial S}{\partial T}=T\left( 
\frac{\partial T}{\partial S}\right) ^{-1}=T\left( \frac{\partial ^{2}m}{%
\partial S^{2}}\right) ^{-1}=\frac{-2S(S-q)}{S-3q}~.  \label{Cq}
\end{equation}%
We can readily observe that the heat capacity diverges along $S=3q$ or
equivalently $q=3m^{2}/4$. No such behaviour occurs for any $q<0$ (for which
we always have $-\infty <C_{q}<0$). In $(m,q)$-coordinates we can express
the heat capacity as 
\begin{equation}
C_{q}=2\Theta \frac{\left( m+\Theta \right) ^{2}}{m-2\Theta }~.
\end{equation}%
Plotted against $\Theta \,$, the heat capacity is shown in Fig \ref{Fig3}. 

The first derivative of the heat capacity at constant charge is given by 
\begin{equation}
\frac{dC_{q}}{d\Theta }=\frac{2\left( m^{2}-\Theta ^{2}\right) \left(
m+4\Theta \right) }{\left( m-2\Theta \right) ^{2}}  \label{Cprime}
\end{equation}%
become singular at $\Theta =m/2$, that is at $q=3m^{2}/4$. In the domain of
its negative values the heat capacity reaches a local maximum at the
Schwarzschild configuration ($\Theta =m$), as can be seen from both Eq. (\ref%
{Cprime}) and Fig \ref{Fig3}.

The thermodynamical interpretation of negative heat capacity is that such a
BH cannot be in a stable equilibrium with an infinite heat reservoir held at 
$T=T_{BH}(m,q)$. For instance, a small thermal fluctuation may transfer some
heat to the BH and make the BH colder, thus making heat transfer even more
efficient. This is the typical behavior of Schwarzschild BHs, which are
unstable with respect to emission of Hawking radiation in empty space and
can be stable only in thermal contact with a finite-volume reservoir. Since
the Universe may be considered as an infinite heat reservoir having the
temperature of the cosmic background radiation, these considerations may be
relevant to the cosmological stability of primordial or near-extremal BHs
that have very low temperature. A near-extremal BH with tidal charge $%
q>3m^{2}/4$ has a positive heat capacity and thus can remain in a stable
equilibrium with an infinite heat reservoir at $T=T_{BH}$.

Nevertheless, Schwarzschild BHs are known to be linearly stable with respect
to perturbations which are nonvanishing on the bifurcation two sphere \cite%
{Wald}.

\subsection{Poincar\'e stability analysis\label{sec1}}

In this subsection we address the issue of microcanonical stability of the
tidal charged BH by using the so-called Poincar\'e method (see \textit{e.g.\ }~%
\cite{ArcTell, Kaburaki, Katz}) which has been used to decide whether
critical phenomena or change of stability occurs in BH systems. The method
is based on analyzing a conjugacy diagram (with a conjugacy parameter, in
our case the inverse temperature, plotted against a control parameter, in
our case the mass) by monitoring the change in the convexity/concavity of
the curve around any occuring "turning point". According to this method
there is a change of stability whenever the concavity/convexity of the curve
changes at a turning point.

\begin{figure}[h]
\includegraphics[scale=0.45]{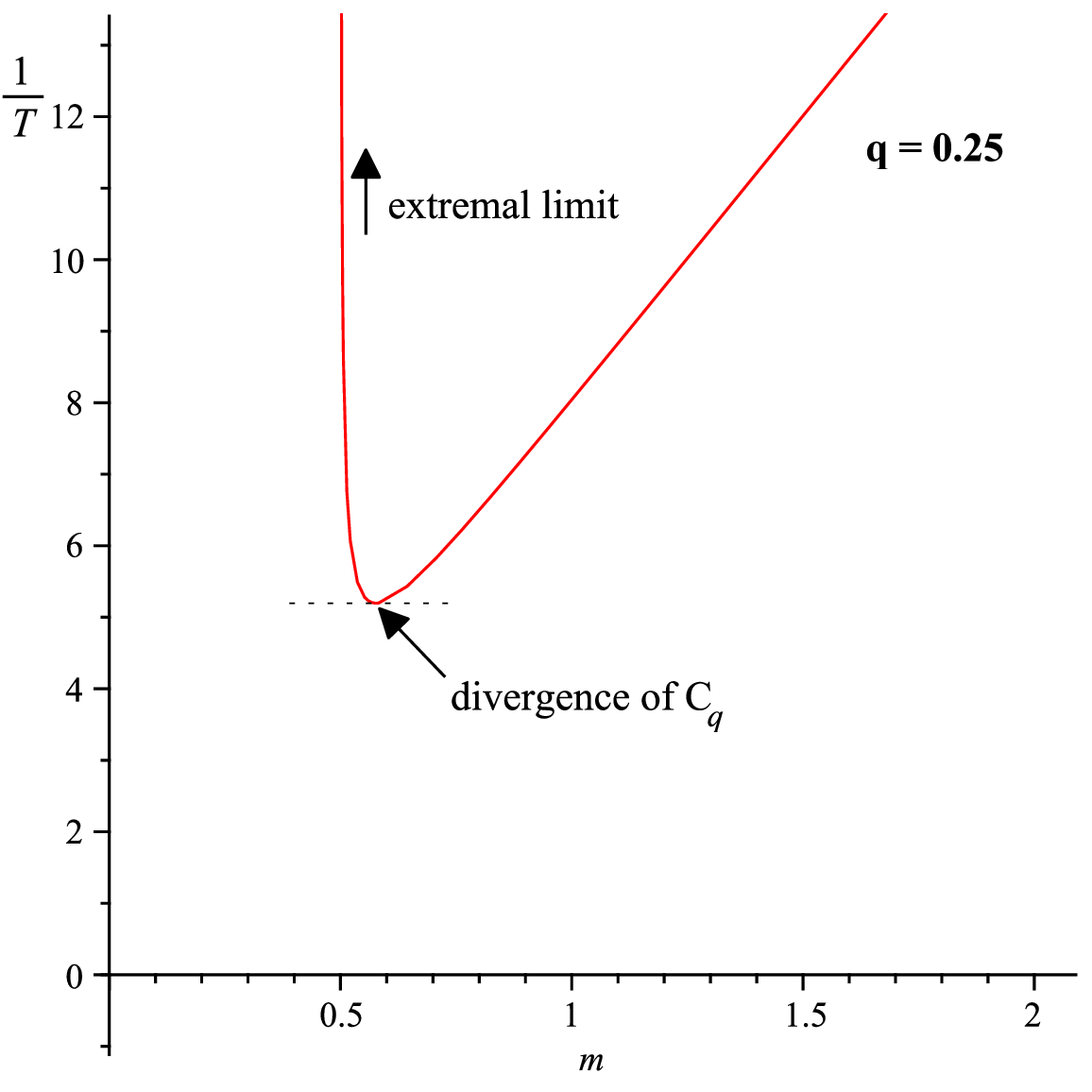} 
\caption{A conjugacy diagram (inverse temperature vs. mass) of the tidal
charged BH for $q=0.25$. The extremal limit is at $m=0.5$ where the curve
goes to infinity. There is no "turning point", so even at $m=0.577$, the
Davies point (where the heat capacity diverges), the stability holds.}
\label{Poincare}
\end{figure}

We plot the conjugacy diagram of the tidal charged BH for $q=0.25\,$\ in Fig %
\ref{Poincare}. In the extremal limit, at $m=0.5$, the curve goes to
infinity. The Davies point where the heat capacity diverges \cite{Davies} is
at $m=0.577$. As the curve has no turning point at all, according to the
Poincar\'e method there is no change of stability in any point, in particular
neither in the Davies point. 

\section{Thermodynamic geometry\label{sec:Thermodynamic-geometry}}

In this section we analyze the information geometries of the tidal charged
BH.

\subsection{The Ruppeiner metric}

The geometry of the tidal charged BH depends on two parameters: $m\,\ $and $%
q $. From the generic definition (\ref{Ruppeiner}) we find the corresponding
Ruppeiner metric as 
\begin{equation}
ds_{R}^{2}=\frac{1}{\Theta ^{3}}\left[ 2\left( m-2\Theta \right) \left(
m+\Theta \right) ^{2}dm^{2}-2(m^{2}-\Theta ^{2})dmdq+\frac{m}{2}dq^{2}\right]
~\text{.}  \label{R}
\end{equation}%
The Ruppeiner curvature scalar is 
\begin{equation}
R=\frac{1}{2\Theta (m+\Theta )}  \label{curvsc}
\end{equation}%
It is readily seen that \textit{the curvature scalar diverges in the
extremal limit} for $q>0$, but stays regular for any $q<0$. It is also worth
to remark that at the Davies point the Ruppeiner metric becomes degenerate
(the coefficient of $dm^{2}$ vanishes). We note here that \textit{for
Reissner-Nordstr\"om BHs the situation was identical}: a singularity in the
heat capacity also emerged \cite{Aman:2003ug} when the metric became
degenerate, and was not accompanied by a phase transition.

\subsection{The Weinhold metric}

By passing to coordinates $\left(m,~\Theta\right)$ in the Ruppeiner metric
and using the conformal relation (\ref{conformal}) as well as the expression
for the temperature (\ref{T}), we obtain the Weinhold metric explicitly as

\begin{equation}
ds_{W}^{2}=Tds_{R}^{2}=\frac{-\left( m+2\Theta \right) dm^{2}-2\Theta
dmd\Theta +md\Theta ^{2}}{\left( m+\Theta \right) ^{2}}~\text{.}
\label{WmTh}
\end{equation}%
This can be further simplified by introducing the new coordinate $r_{+}$
replacing $\Theta $:%
\begin{equation}
ds_{W}^{2}=\frac{dr_{+}}{r_{+}}\left( m\frac{dr_{+}}{r_{+}}-2dm\right) ~%
\text{,}  \label{Wmr+}
\end{equation}%
then passing to $\left( Z=\log r_{+},~W=\log \left( r_{+}/m^{2}\right)
\right) $ we find%
\begin{equation}
ds_{W}^{2}=mdZdW\,,
\end{equation}%
with 
\begin{equation}
m=\exp \left( \frac{Z-W}{2}\right) ~.
\end{equation}%
In the coordinates $\left( U_{+}=2\exp \left( Z/2\right) ,~U_{-}=2\exp
\left( -W/2\right) \right) $ the Weinhold metric becomes \textit{manifestly
flat}, $ds_{W}^{2}=-dU_{+}dU_{-}$. One can also introduce Minkowskian
coordinates as $U_{\pm }=X\pm Y$, finding $ds_{W}^{2}=-dX^{2}+dY^{2}$. The
sequence of coordinate transformations leading to this result can be
summarized as%
\begin{eqnarray}
X &=&\sqrt{r_{+}}+\frac{m}{\sqrt{r_{+}}}~,  \notag \\
Y &=&\sqrt{r_{+}}-\frac{m}{\sqrt{r_{+}}}~.  \label{transf}
\end{eqnarray}%
The inverse transformation is%
\begin{eqnarray}
4r_{+} &=&\left( X+Y\right) ^{2}~,  \notag \\
4m &=&X^{2}-Y^{2}~.  \label{inverse}
\end{eqnarray}

\subsection{Direct derivation of the Weinhold metric}

As a further consistency check of our calculations we express from Eq. (\ref%
{entropy}) the mass $m$ of the tidal charged BH in $\left( S,q\right) $
coordinates, obtaining 
\begin{equation}
m=\frac{\sqrt{S}}{2}\left( 1+\frac{q}{S}\right)
\end{equation}%
$\allowbreak $This allows to calculate the Weinhold metric directly from its
definition:%
\begin{equation}
ds_{W}^{2}=\frac{3q-S}{8S^{5/2}}dS^{2}-\frac{1}{2S^{3/2}}dSdq~.  \label{WSq}
\end{equation}%
It can easily be checked that the above metric is flat. When $S=3q$ (in the
Davies point, where the heat capacity diverges) the Weinhold metric is also
degenerate. By performing a transformation to $\left( m,\Theta \right) $
coordinates, we recover the form (\ref{WmTh}) of the Weinhold metric.$%
\allowbreak $

\subsection{The global structure of the Ruppeiner geometry}

The expression of the temperature in the $\left( X,~Y\right) $ coordinates
is 
\begin{equation}
T=\frac{r_{+}-m}{2r_{+}^{2}}=\frac{4Y}{\left( X+Y\right) ^{3}}~,  \label{Tq}
\end{equation}%
which leads to the manifestly conformally flat form of the Ruppeiner metric:%
\begin{equation}
ds_{R}^{2}=\frac{\left( X+Y\right) ^{3}}{4Y}\left( -dX^{2}+dY^{2}\right) ~.
\end{equation}

\begin{figure}[tbp]
\includegraphics[width=9cm]{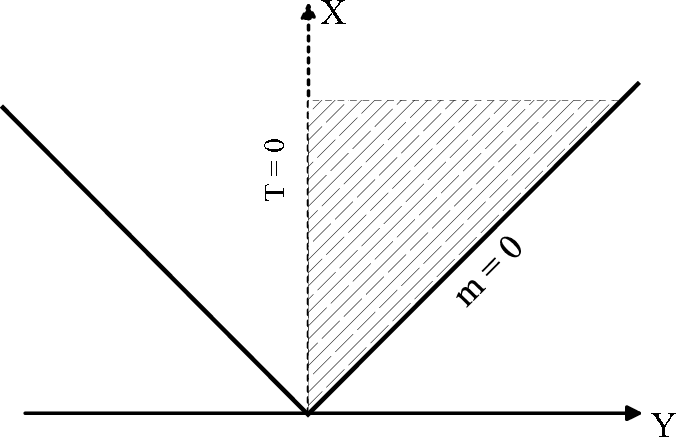} 
\caption{A state space plot of the tidal charged BH embedded in the flat
Minkowskian parameter space. Note that the thermodynamic light cone (TLC)
describes $m=0$ and the wedge fills the right half of the TLC with the TLC
itself excluded. The vertical axis represents the extremal limit in which $%
T=0$.}
\label{fig4}
\end{figure}
Note that the domain of the original Ruppeiner coordinates is 
\begin{equation}
m\in\left(0,~\infty\right),~q\in\left(-\infty,~m^{2}\right)\,.
\end{equation}
The corresponding ranges of the variables $\Theta,r_{+}$ are $%
\Theta\geq0,~r_{+}\geq0$; the Minkowskian coordinates defined by (\ref%
{transf}) have the range $X>Y\geq0.$ Thus the state space is equivalent to
the right half of the interior of the future light cone of a Minkowski
plane, with the vertical boundary included but the light-like boundary
excluded. (The light cone describes $m=0$ states as can be seen from $%
4m=X^{2}-Y^{2}$, which for $q\geq0$ does not correspond to BH metrics.) The
extremal states are located at $\left(X=2\sqrt{m}>0,~Y=0\right)$, i.e.~on
the positive half of the time-like coordinate axis (the vertical boundary).
This can be also seen by writing the curvature scalar (\ref{curvsc}) of the
Ruppeiner metric in the $\left(X,Y\right)$ coordinates:%
\begin{equation}
R=\frac{1}{2\left(r_{+}-m\right)r_{+}}=\frac{4}{Y\left(X+Y\right)^{3}}~.
\end{equation}

We also remark that passing to the $\left( X,~Y\right) $ coordinates by the
transformation (\ref{inverse}) induces a degeneracy. For each pair of
coordinates $\left( m,~q\right) $, as well as for $\left( m,\Theta \right) $
or $\left( m,~r_{+}\right) $, we can associate any of the combinations $%
\left( \pm X\text{, }\pm Y\right) $, with $X,~Y$ defined by Eq. (\ref{transf}%
). Therefore the light cone of the Minkowski plane provides a four-fold
coverage of the original state space. This is similar to the introduction of
the well-known Kruskal coordinates for the Schwarzschild geometry: Kruskal
coordinates cover four patches in the Kruskal-Szekeres diagram, while the
original coordinates cover only one patch. See Fig.~\ref{fig4}.

\section{Comparison and physical interpretation of thermodynamic properties
for the tidal charged and Reissner-Nordstr\"om BHs\label{sec:stable}}

After the generic consideration on the tidal charged BH thermodynamics made
in Section \ref{sec:Thermodynamic-considerations} and the detailed analysis
of the related information geometries in Section \ref%
{sec:Thermodynamic-geometry}, we are able to compare the tidal charged BH
and the general relativistic Reissner-Nordstr\"om BH from a thermodynamic
point of view.

The entropy of a Reissner-Nordstr\"om BH still obeys Eq. (\ref{entropy}), but
with $\Theta _{RN}=\sqrt{m^{2}-Q^{2}}$. Its Ruppeiner geometry is flat,
while the curvature scalar of its Weinhold geometry in ($S_{RN},Q$)
coordinates is \cite{Aman:2003ug}%
\begin{equation}
R_{W}^{RN}=\frac{2S_{RN}^{3/2}}{\left( S_{RN}-Q^{2}\right) ^{2}}~.
\end{equation}%
We can rewrite this in the ($m,\Theta _{RN}$) coordinates as 
\begin{equation}
R_{W}^{RN}=\frac{m+\Theta _{RN}}{2\Theta _{RN}^{2}}~.
\end{equation}%
The temperature of the Reissner-Nordstr\"om BH is \cite{Aman:2003ug}:%
\begin{equation}
T_{RN}=\frac{1}{4S_{RN}^{1/2}}\left( 1-\frac{Q^{2}}{S_{RN}}\right) ~.
\end{equation}%
In the ($m,\Theta _{RN}$) coordinates it reads%
\begin{equation}
T_{RN}=\frac{\Theta _{RN}}{2\left( m+\Theta _{RN}\right) ^{2}}~,
\end{equation}%
which has the same functional form as Eq. (\ref{T}).

Summarizing (a) the temperatures of the general relativistic Reissner-Nordstr\"om BH and tidal charged BH have the same functional form, and (b) both
information geometries are regular for both of the BHs, with the sole
exception of the extremal configurations, where the Ruppeiner metric for the
tidal charged BH and the Weinhold metric for the Reissner-Nordstr\"om BH
diverge. The other two information geometries are flat. The comparison of
the information geometries of the two BHs is presented in Table \ref{RNtidal}%
. 
\begin{table}[t]
\caption{Ruppeiner and Weinhold information geometries for the
Reissner-Nordstr\"om and tidal charged black holes. As the state space for the
information geometries is two-dimensional, the respective geometries are
fully characterized by the curvature scalars (second and third columns). The
conformal factor relating the Weinhold and Ruppeiner geometries is the
temperature of the respective black holes (fourth column). }
\label{RNtidal}
\begin{center}
\begin{tabular}{c|c|c|c}
& Weinhold & Ruppeiner & temperature \\ \hline
Reissner-Nordstr\"om BH & $R_{W}^{RN}=\frac{m+\Theta _{RN}}{2\Theta _{RN}^{2}}$
& $R_{R}^{RN}=0$ & $T_{RN}=\frac{\Theta _{RN}}{2\left( m+\Theta _{RN}\right)
^{2}}$ \\ 
tidal charged BH & $R_{W}^{tidal}=0$ & $R_{R}^{tidal}=\frac{1}{2\Theta
\left( m+\Theta \right) }$ & $T=\frac{\Theta }{2\left( m+\Theta \right) ^{2}}$
\end{tabular}
\end{center}
\end{table}

In order to see how this fits into a more generic context, we reproduce here
a comparative table \ref{compare} of the information geometries for various
BHs, given first in \cite{Narit-thesis}. 
\begin{table}[h]
\caption{Comparison of information geometries for various black holes.}
\label{compare}
\begin{center}
\begin{tabular}{|c|l|l|l|}
\hline\hline
{\small \textbf{Spacetime dimension}} & \textbf{Black hole family} & \textbf{%
Ruppeiner} & \textbf{Weinhold} \\ \hline
$d=2$ & (1+1) RN like BH (generic) & Curved & Curved \\ \cline{2-4}
& (1+1) reduced RN BH & Flat & Curved \\ \cline{2-4}
& (1+1) CS like BH (generic) & Curved & Flat \\ \hline
$d=3$ & (2+1) BTZ & Flat & Curved \\ \cline{2-4}
& (2+1) BTZ (Chern-Simons) & Flat & Curved \\ \cline{2-4}
& (2+1) BTZ (Log corrections) & Curved & Curved \\ \hline
$d=4$ & RN & Flat & Curved \\ \cline{2-4}
& Kerr & Curved & Flat \\ \cline{2-4}
& Kerr-Newman & Curved & Curved \\ \cline{2-4}
& Braneworld (tidal charged) & Curved & Flat \\ \cline{2-4}
& Dilaton & Flat & Curved \\ \hline
$d=5$ & Kerr & Curved & Flat \\ \cline{2-4}
& double-spin Kerr & Curved & Curved \\ \cline{2-4}
& RN & Flat & Curved \\ \cline{2-4}
& Black ring & Curved & Flat \\ \hline
\text{any} $d$ & Kerr & Curved & Flat \\ \cline{2-4}
& RN & Flat & Curved \\ \hline\hline
\end{tabular}%
\end{center}
\end{table}

The Ruppeiner metric does not have singularities for any of these BHs, with
the exception of the extremal tidal charged BH. Therefore, based on generic
result that instability are accompanied by singularities in the
Ruppeiner geometry \cite{Aman:2003ug}, \cite{Aman:2006kd}, \cite{ArcTell}\
one would not expect phase transitions for these BHs, except perhaps the
extremal tidal charged configuration. Nevertheless, extremal BHs have zero
temperature, therefore this is a special case which will be addressed later.

\begin{figure}[h]
\includegraphics[scale=0.45]{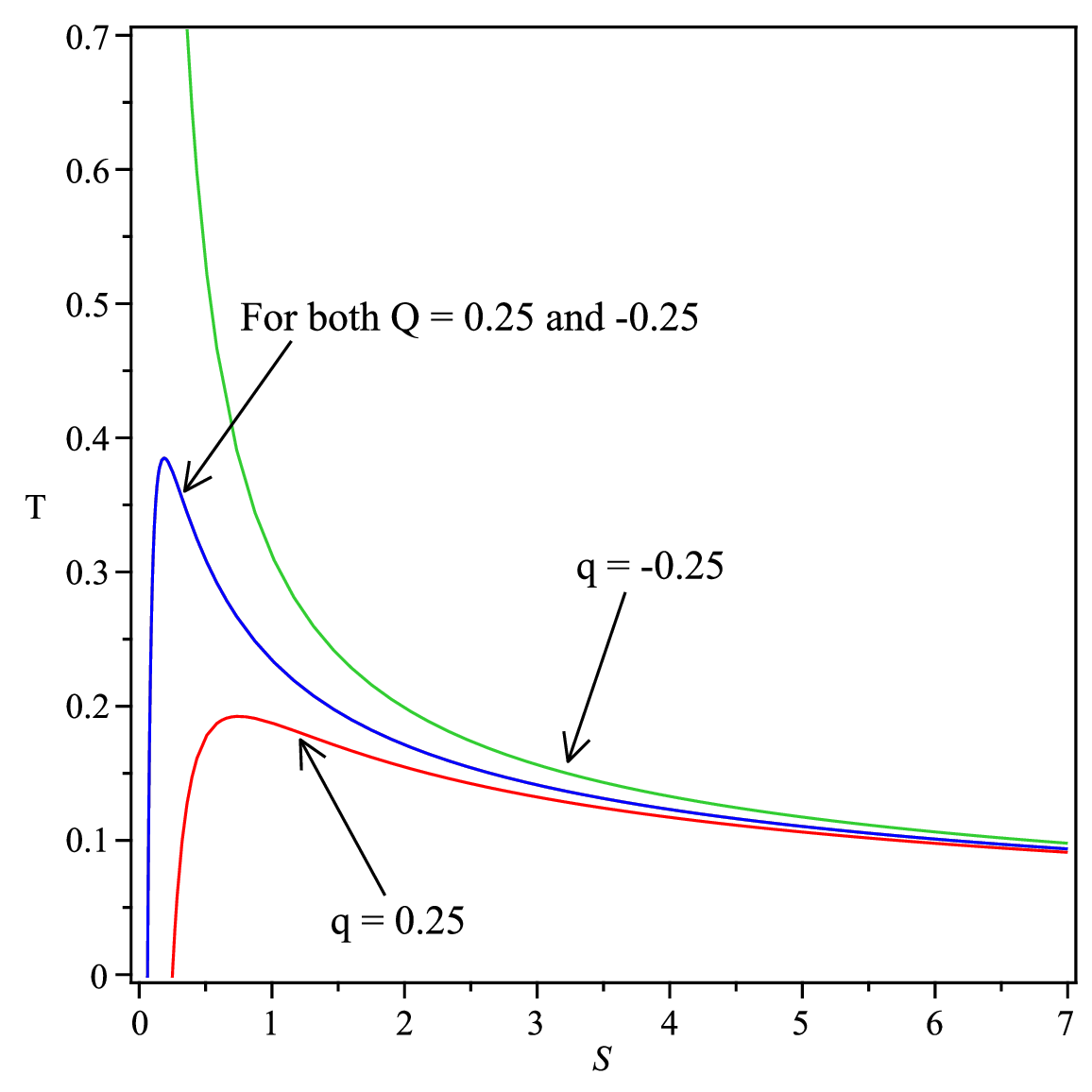} 
\caption{The ($T,S$) diagram for the Reissner-Nordstr\"om and tidal charged
black holes. Negative tidal charged black holes are singled out in having no
Davies point.}
\label{TS}
\end{figure}

The heat capacity diverges at $q=3m^{2}/4$ for tidal charged BHs, and at $%
Q=3^{1/2}m/2$ for the Reissner-Nordstr\"om BH \cite{Aman:2006kd}. These Davies
points also show up on the ($T,S$) diagram represented in Fig. \ref{TS}.
These divergences share a similar interpretation. While such a divergence is
characteristic to second order phase transitions in ordinary thermodynamics,
for gravitating systems the situation is changed. In both cases mentioned
above, besides diverging, the heat capacity also changes sign in the
respective points. According to Sorkin a microcanonical instability would
occur only if the heat capacity changes sign through zero \cite{Sorkin},
which is not the case for either of the Reissner-Nordstr\"om or tidal charged
BHs. This is in full agreement with the regularity of the Ruppeiner metric
in the respective points.

Katz et al. have argued \cite{Katz}, that instabilities do not necessarily
come together with divergences of the heat capacity. Their argument was
based on the Poincar\'e stability analysis. The Poincar\'e stability analyses of
both the Reissner-Nordstr\"om and tidal charged BHs (the latter in this paper)
have shown no instabilities at the diverging heat capacity state-space
configurations. Moreover, the Poincar\'e stability analysis does not indicate
any instability at the extremal configuration either, see Fig. \ref{Poincare}%
, thus the extremal tidal charged BH does not exhibit a phase transition
there. Based on the idea that the extremal limit of various black hole families might themselves be regarded as critical points~\cite{extremal-critical}, 
the Reissner-Nordstr\"om extremal point ought to represent some type of phase transition. Since the Ruppeiner metric 
for the Reissner-Nordstr\"om  black hole is flat and a Poincar\'e stability analysis \cite{Katz} shows no sign of instability 
at the extremal point, therefore we conclude that there are no phase transitions associated with it. This serves as a basis for our analysis of the tidal charged 
black hole in the extremal limit. 

\begin{figure}[h]
\includegraphics[scale=0.45]{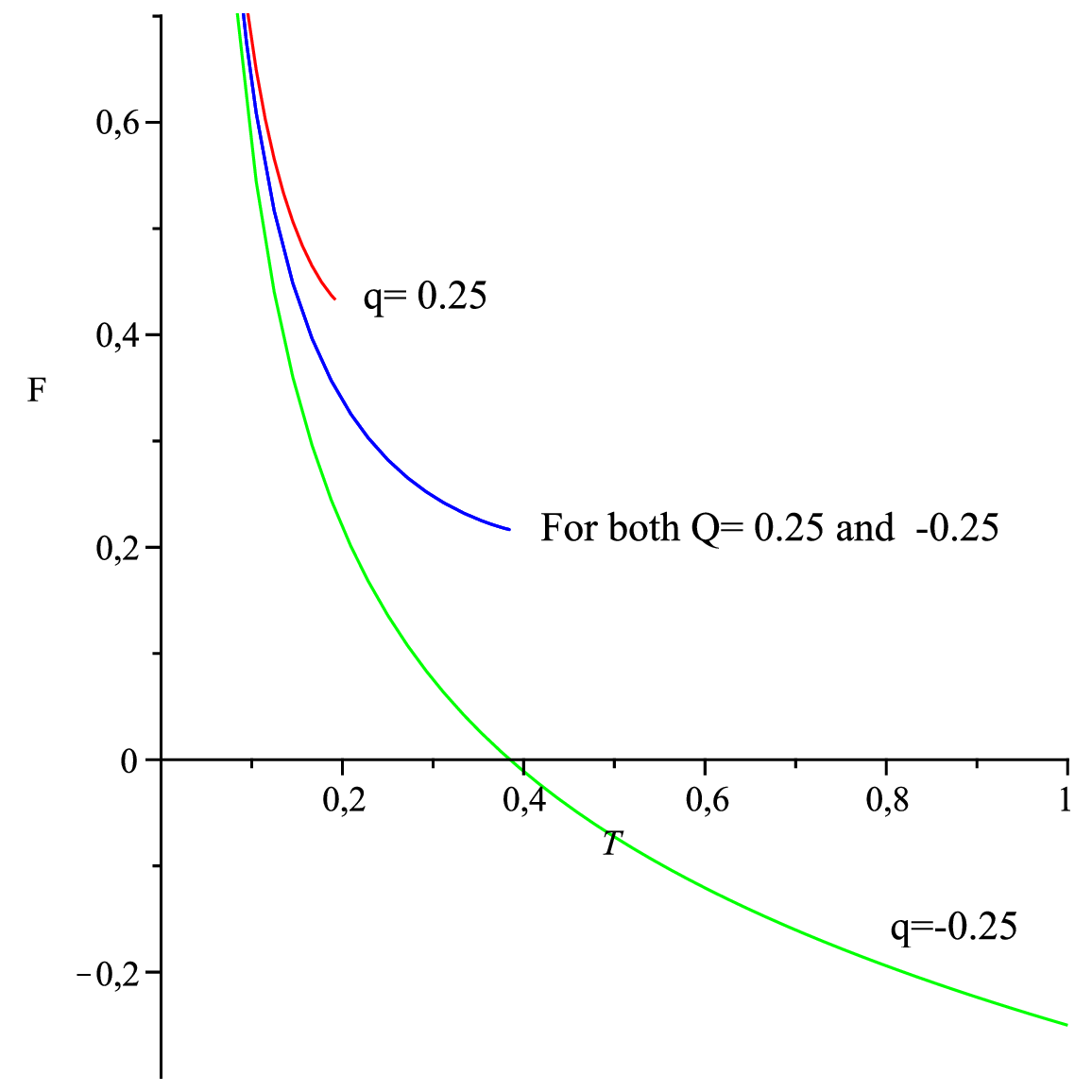} 
\caption{The free energy vs. temperature for the Reissner-Nordstr\"om and
tidal charged black holes. With the exception of the negative tidal charged
black hole, in all other cases the plots terminate at a finite temperature,
indicating the existence of Davies points, as discussed in \protect\cite{La}%
. }
\label{free}
\end{figure}

We conclude this section with comments on the negative tidal charge regime.
As can be seen from Fig. \ref{TS}, this range has no Davies point. This is
also supported by the analysis of the free energy $F=m-TS$, plotted for both
the Reissner-Nordstr\"om and tidal charged cases in Fig. \ref{free}. Whereas
the respective curves end at the Davies point for any value of the electric
charge, in accordance with Ref. \cite{La} (including zero charge, the
Schwarzschild case), also for any positive value of the tidal charge, the
curve belonging to the negative tidal charge case continues towards
infinity, signalling that not even Davies points can exist for negative
tidal charge. 

The main message of our analysis of this section is that \textit{neither the
Reissner-Nordstr\"om, nor the tidal charged BHs allow for phase transitions}
at any of their parameter values.

\section{Hawking limit for tidal charged BHs\label{sec:limit}}

\begin{figure}[h]
\includegraphics[scale=0.45]{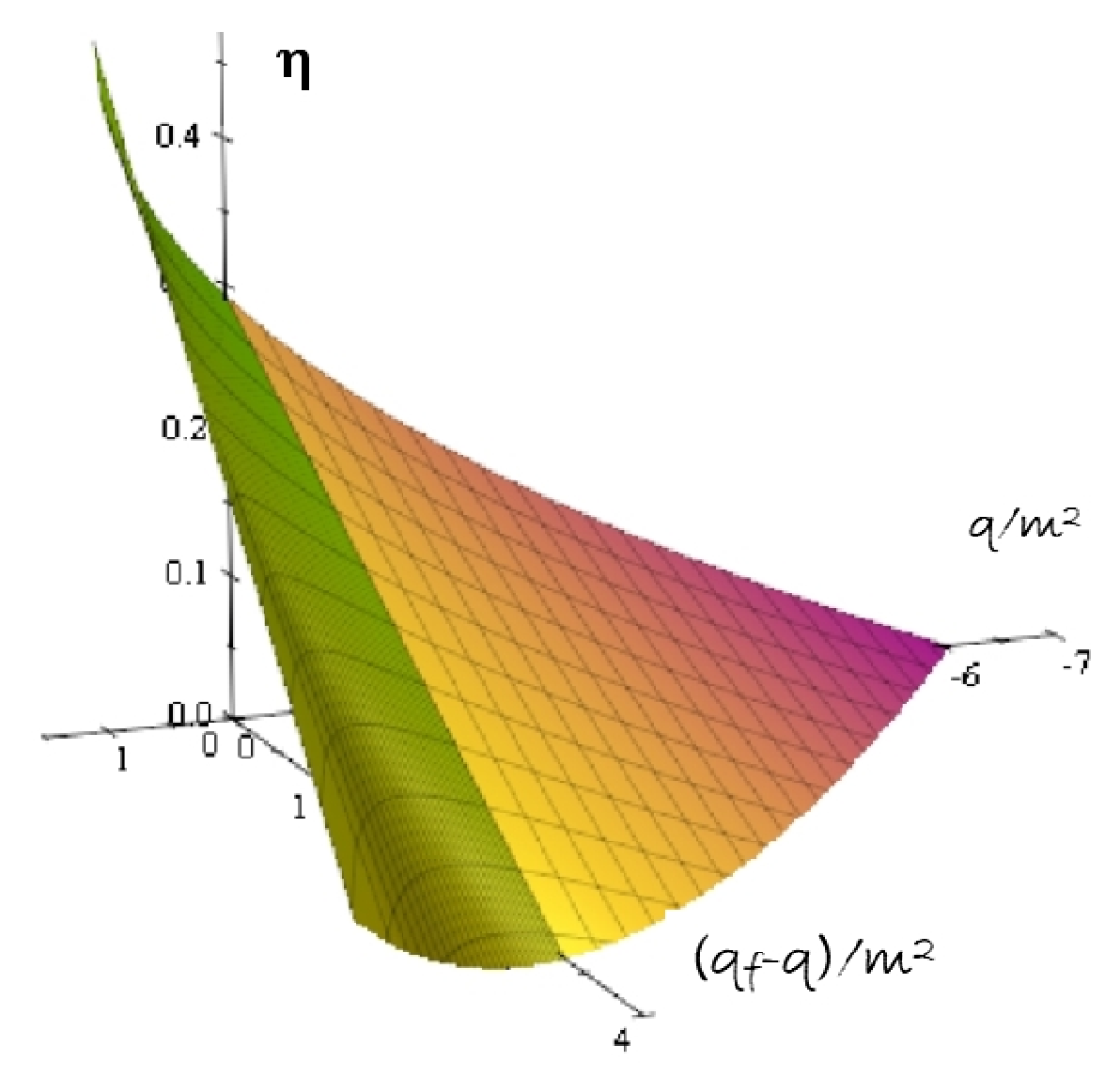} 
\caption{The efficiency limit of mass conversion into radiation for equal
mass, equal tidal charged black hole collissions restricts both the values
of $q/m^{2}$ and the possible value of the tidal charge of the final black
hole. }
\label{efficiency_limit}
\end{figure}

In a seminal paper \cite{Hawking}, Hawking has derived upper limits for the
efficiency 
\begin{equation}
\eta =1-\frac{m_{f}}{m_{1}+m_{2}}
\end{equation}%
of mass conversion into gravitational radiation, when BHs merge. Here $m_{1,2}$
are the masses of the individual BHs and $m_{f}$ is the mass of the final
product. The merger of two equal mass, non-rotating BHs leads to an upper
efficiency limit of $1-2^{-1/2}=\allowbreak 29.3~\%$. With rotation
included, the efficiency limit can increase up to $50\%$. The basic argument
is provided by the second law of BH thermodynamics $S_{f}\geq S_{1}+S_{2}$.
Employing Eq. (\ref{entropydef}) for the tidal charged BHs this leads to $%
\left( m_{f}+\Theta _{f}\right) ^{2}\geq \left( m_{1}+\Theta _{1}\right)
^{2}+\left( m_{2}+\Theta _{2}\right) ^{2}$, thus%
\begin{equation}
m_{f}\geq \frac{m_{1}^{2}\left( 1+\sqrt{1-\frac{q_{1}}{m_{1}^{2}}}\right)
^{2}+m_{2}^{2}\left( 1+\sqrt{1-\frac{q_{2}}{m_{2}^{2}}}\right) ^{2}+q_{f}}{2%
\left[ m_{1}^{2}\left( 1+\sqrt{1-\frac{q_{1}}{m_{1}^{2}}}\right)
^{2}+m_{2}^{2}\left( 1+\sqrt{1-\frac{q_{2}}{m_{2}^{2}}}\right) ^{2}\right]
^{1/2}}~.
\end{equation}%
For identical BHs $m_{2}=m_{1}\equiv m$ and $q_{1}=q_{2}\equiv q<q_{f}$ (the
tidal charge being an extensive parameter, as assumed throughout this
paper), the efficiency limit becomes%
\begin{equation}
\eta \leq 1-\frac{2+2\sqrt{1-\frac{q}{m^{2}}}-\frac{q}{2m^{2}}+\frac{q_{f}-q%
}{2m^{2}}}{2^{3/2}\left[ 2+2\sqrt{1-\frac{q}{m^{2}}}-\frac{q}{m^{2}}\right]
^{1/2}}~.
\end{equation}%
This is represented on Fig. \ref{efficiency_limit}, as function of $%
q/m^{2}\in (-\infty ,1]$ and $\left( q_{f}-q\right) /m^{2}>0$. We can derive
a couple of important restrictions on the physically allowed range of the
tidal charged BH parameters from this plot:

(a) The ratio $q/m^{2}$ is bounded from below by a value $\in \left[ -6,0%
\right] $ depending on $\left( q_{f}-q\right) /m^{2}$.

(b) The quantity $\left( q_{f}-q\right) /m^{2}$, a measure of the
extensiveness of the tidal charge is also bounded.

Both constraints on the state space parameters come from requiring that the
radiated mass cannot be more than the sum of the initial masses. To our
knowledge these are the first constraints on the possible range of the tidal
charge, derived in the literature. 

\section{Concluding remarks}

We have studied the thermodynamics and thermodynamic geometries (both
Ruppeiner and Weinhold) of the tidal charged brane BH. While the
thermodynamic state space of the Reissner-Nordstr\"om BH is a Rindler wedge
embedded in a Minkowski parameter space, the state space of the tidal
charged BH is the right half of the interior of the future light cone of the
Minkowski plane, with the vertical boundary included but the light-like
boundary excluded. The light cone of the Minkowski plane provides a
four-fold coverage of this state space, similarly to the four-fold covering
of the curvature coordinates for a Schwarzschild BH by Kruskal coordinates.
The geodesics of the information geometries characterize the time for 
\textit{quasistatic relaxation} and \textit{minimum dissipation},
respectively.

Whilst from the point of view of gravitational curvature the $q>0$ case is
the Reissner-Nordstr\"om geometry with electric charge $Q=\sqrt{q}$, a
negative tidal charge has no correspondence in general relativity. Along the
well-known property that such a negative tidal charge strengthens the
gravitational field of the BH and as such contributes to the localization of
gravity on the brane, our analysis has explicitly shown that $q<0$ is also 
\textit{thermodynamically preferred}, as it comes with a higher entropy.

For $q>0$ we have found both similarities and differences as compared to the
Reissner-Nordstr\"om BH. The Ruppeiner geometry having a positive Ricci
curvature, as opposed to the flat Ruppeiner metric for the Reissner-Nordstr\"om BH \cite{Aman:2003ug}, indicates that \textit{the underlying statistical
models for the two BHs could be different}.

Despite the heat capacity of the tidal charged BH diverging on a set of
measure zero of the parameter space (at $q=3m^{2}/4$, the Davies point),
both a Poincar\'e stability analysis and the regularity of the Ruppeiner
metric shows no indication of phase transition at that point, similarly as
for the Reissner-Nordstr\"om BH and the Kerr BH \cite{ArcTell}, where (despite
the change of sign of the heat capacity through infinite values at the
Davies point) the BHs remains stable. Similarly for some other BHs, the induced
Ruppeiner geometry of the tidal charged BH is singular in the extremal
limit, but Poincar\'e stability analysis once again disrules the possibility
of phase transition there. Therefore we have shown that \textit{the tidal
charged BH is stable for the full parameter range}.

Starting from the Hawking limit for the efficiency of gravitational
radiation in a BH merger process, we have derived here for the first time 
\textit{constraints on the tidal charge range}.

\section{Acknowledgments}

LG wishes to acknowledge discussions with Shinji Mukohyama on the relation
of the entropies of higher dimensional black objects and their lower
dimensional sections. He was succesively supported by the Hungarian
Scientific Research Fund (OTKA) grant no. 69036, a London South Bank
University Research Opportunities Fund Award, the Polnyi and Sun Programs
of the Hungarian National Office for Research and Technology (NKTH), by
Collegium Budapest, finally by COST Action MP0905 "Black Holes in a Violent
Universe". NP would like to thank Roberto Emparan for insightful discussions
and useful comments, and also wishes to acknowledge the hospitality of
Departament de Fsica Fonamental at Universitat de Barcelona during his
research visit. He was supported by Doktorandtjnst of Stockholm University
during the early stage of this work and is grateful for scholarships from
Helge Ax:son Johnsons stiftelse and K \& A Wallenbergs stiftelse. NP would also like
to acknowledge the kind hospitality of the Department of Theoretical Physics and the Department of Experimental Physics 
at the University of Szeged, Hungary during his recent visit funded by the Royal Swedish Academy of Sciences (KVA).

\appendix

\section{The general relation between the Ruppeiner and Weinhold metrics 
\label{sec:conformal}}

In this Appendix we present a concise derivation of the generic statement
that the Ruppeiner and Weinhold metrics (\ref{Ruppeiner}) and (\ref{Weinhold}%
) are conformally related, with the temperature (\ref{temperature}) as the
conformal factor.

Let us consider a scalar function\footnote{%
The function $m$ is generic at this stage, but will be identified with the
mass at the end of the Appendix.} $m\left(x\right)$ of the variables $%
x\equiv\left\{ x^{a}\mid a=\overline{1,n+1},~n\in N\right\} $. We will also
consider another coordinate system $y\equiv\left\{ y^{c}\mid c=\overline{%
1,n+1},~n\in N\right\} $ in the same configuration space. A series expansion
of $\delta m\equiv m(x+\delta x)-m(x)$ about an arbitrary point gives%
\begin{eqnarray}
\delta m\left(x\right) & = & \frac{\partial m}{\partial x^{a}}\delta x^{a}+%
\frac{1}{2}\frac{\partial^{2}m}{\partial x^{a}\partial x^{b}}\delta
x^{a}\delta x^{b}+\mathcal{O}\left(\delta x\right)^{3}~,  \label{1} \\
\delta m\left(y\right) & = & \frac{\partial m}{\partial y^{c}}\delta y^{c}+%
\frac{1}{2}\frac{\partial^{2}m}{\partial y^{c}\partial y^{d}}\delta
y^{c}\delta y^{d}+\mathcal{O}\left(\delta y\right)^{3}~.  \label{2}
\end{eqnarray}
It is understood that all derivatives are evaluated at the same point $x_{0}$%
, and that $y_{0}$ and $\delta y_{0}$ represent $x_{0}$ and $\delta x_{0}$
expressed in the coordinates $\left\{ y^{c}\right\} $. In particular, we
have a similar expansion 
\begin{equation}
\delta x^{a}\left(y\right)=\frac{\partial x^{a}}{\partial y^{c}}\delta y^{c}+%
\frac{1}{2}\frac{\partial^{2}x^{a}}{\partial y^{c}\partial y^{d}}\delta
y^{c}\delta y^{d}+\mathcal{O}\left(\delta y\right)^{3}~.  \label{d3}
\end{equation}
Let us now derive a relationship between $\partial^{2}m/\partial x^{a}{%
\partial x}^{b}$ and $\partial^{2}m/\partial y^{c}\partial y^{d}$. By
inserting Eq.~(\ref{d3}) into Eq.~(\ref{1}), we obtain a second expression
for $\delta m$ as a function of the coordinates $y$; that expression should
coincide with Eq.~(\ref{2}) order by order. By identifying the first-order
and second-order contributions with the respective quantities of Eq.~(\ref{2}%
) we obtain:%
\begin{eqnarray}
\frac{\partial m\left(y\right)}{\partial y^{c}} & = & \frac{\partial
m\left(x\right)}{\partial x^{a}}\frac{\partial x^{a}\left(y\right)}{\partial
y^{c}}~,  \label{4} \\
\frac{\partial^{2}m\left(y\right)}{\partial y^{c}\partial y^{d}} & = & \frac{%
\partial^{2}m\left(x\right)}{\partial x^{a}\partial x^{b}}\frac{\partial
x^{a}\left(y\right)}{\partial y^{c}}\frac{\partial x^{b}\left(y\right)}{%
\partial y^{d}}+\frac{\partial m\left(x\right)}{\partial x^{a}}\frac{%
\partial^{2}x^{a}\left(y\right)}{\partial y^{c}\partial y^{d}}~.
\label{cond}
\end{eqnarray}
While Eq.~(\ref{4}) is simply the chain rule, Eq.~(\ref{cond}) is the
identity that will eventually yield the desired result.

In order to achieve this, we specify the coordinates as follows,%
\begin{eqnarray}
x &=&\left\{ q^{1},...,q^{n},S\right\} ~,  \label{5} \\
y &=&\left\{ q^{1},...,q^{n},m\right\} ~.  \label{6}
\end{eqnarray}%
In other words, the coordinates $y^{i}$ are the same as $x^{i}$ except for
the last coordinate $y^{n+1}$, which is chosen as the function $m(x)$
itself. It then follows that 
\begin{equation}
\frac{\partial m(y)}{\partial y^{i}}=\delta _{i}^{n+1}\,,
\end{equation}%
thus the left hand side of Eq.~(\ref{cond}) identically vanishes. It is also
straightforward to see that%
\begin{equation}
\frac{\partial ^{2}x^{a}\left( y\right) }{\partial y^{c}\partial y^{d}}=%
\frac{\partial }{\partial y^{d}}\left[ \delta _{i}^{a}\delta _{c}^{i}+\delta
_{n+1}^{a}\frac{\partial S\left( y\right) }{\partial y^{c}}\right] =\delta
_{n+1}^{a}\frac{\partial ^{2}S\left( y\right) }{\partial y^{c}\partial y^{d}}%
\,.  \label{7}
\end{equation}%
Thus Eq.~(\ref{cond}) is reduced to the identity 
\begin{equation}
\frac{\partial ^{2}m\left( x\right) }{\partial x^{a}\partial x^{b}}\frac{%
\partial x^{a}\left( y\right) }{\partial y^{c}}\frac{\partial x^{b}\left(
y\right) }{\partial y^{d}}=-\frac{\partial m\left( x\right) }{\partial S}%
\frac{\partial ^{2}S\left( y\right) }{\partial y^{c}\partial y^{d}}~.
\label{8}
\end{equation}%
By recalling the definitions (\ref{Ruppeiner}), (\ref{Weinhold}) and (\ref%
{temperature}) of the Ruppeiner metric, Weinhold metric and temperature,
respectively, by having in mind that the set of coordinates $x$ and $y$ are
given by Eqs. (\ref{5})-(\ref{6}), we have recovered the desired relation%
\begin{equation}
g_{ab}^{W}\left( x\right) \frac{\partial x^{a}}{\partial y^{c}}\frac{%
\partial x^{b}}{\partial y^{d}}=Tg_{cd}^{R}\left( y\right) ~,
\label{relation1}
\end{equation}%
where the left hand side represents the Weinhold metric transformed to the $%
y $ coordinates.

\section{The geodesics of the thermodynamic metrics\label{sec:geodesics}}

The interpretation of the thermodynamic metrics is the following. A Weinhold
geodesic (the path of least Weinhold length) in the parameter space
corresponds to a quasistatic process that, in a well-defined sense, \textit{%
optimizes the {}\textquotedblleft thermodynamical time\textquotedblright },
which roughly means the time necessary for \textit{quasistatic relaxation}
during the process~\cite{Diosi1996}. Since in the present case the Weinhold
metric is flat, its geodesics are straight lines both in the Minkowskian
coordinates $\left( X,Y\right) $ and in the null coordinates $U_{\pm }=X\pm
Y $:%
\begin{equation}
U_{+}=\alpha U_{-}+\beta ~,  \label{gW}
\end{equation}%
with $\alpha ,~\beta $ constants.

On the other hand, the {}\textquotedblleft length\textquotedblright\ of a
process computed in the Ruppeiner metric describes the \textit{minimum
dissipation} (increase of entropy) that is unavoidable in a quasistatic
process~\cite{Nulton1985}. The geodesic equations for the Ruppeiner metric
are the simplest when written in the conformally Minkowskian null
coordinates $U_{\pm }$:%
\begin{eqnarray}
\frac{d^{2}U_{+}}{d\lambda ^{2}}+\frac{2U_{+}-3U_{-}}{U_{+}\left(
U_{+}-U_{-}\right) }\left( \frac{dU_{+}}{d\lambda }\right) ^{2} &=&0~, 
\notag \\
\frac{d^{2}U_{-}}{d\lambda ^{2}}+\frac{1}{U_{+}-U_{-}}\left( \frac{dU_{-}}{%
d\lambda }\right) ^{2} &=&0~,  \label{gR}
\end{eqnarray}%
with $\lambda $ an affine parameter.

\end{document}